# A High Energy-Efficiency Multi-core Neuromorphic Architecture for Deep SNN Training

Mingjing Li[†,1], Huihui Zhou[†,*,1], Xiaofeng Xu[†,1], Zhiwei Zhong[†,1], Puli Quan[1], Xueke Zhu[1], Yanyu Lin[1], Wenjie Lin[1], Hongyu Guo[1], Junchao Zhang[1], Yunhao Ma[1,5], Wei Wang[1], Qingyan Meng[1], Zhengyu Ma[1], Guoqi Li[*,1,4], Xiaoxin Cui[*,1,2], Yonghong Tian[*,1,2,3]

[†] These authors contributed equally
[*] Corresponding authors, emails: zhouhh@pcl.ac.cn, guoqi.li@ia.ac.cn, cuixx@pku.edu.cn, yhtian@pku.edu.cn
1. PengCheng Laboratory, Shenzhen, China
2. School of Computer Science, Peking University, Beijing, China
3. School of Electronic and Computer Engineering, Shenzhen Graduate School, Peking University, Shenzhen, China
4. Institute of Automation, Chinese Academy of Sciences, Beijing, China
5. Southern University of Science and Technology, Shenzhen, China

**Abstract:** There is a growing necessity for edge training to adapt to dynamically changing environment. Neuromorphic computing represents a significant pathway for high-efficiency intelligent computation in energy-constrained edges, but existing neuromorphic architectures lack the ability of directly training spiking neural networks (SNNs) based on backpropagation. We develop a multi-core neuromorphic architecture with Feedforward-Propagation, Back-Propagation, and Weight-Gradient engines in each core, supporting highly efficient parallel computing at both the engine and core levels. It combines various data flows and sparse computation optimization by fully leveraging the sparsity in SNN training, obtaining a high energy efficiency of 1.05TFLOPS/W@ FP16 @ 28nm, 55 ~ 85% reduction of DRAM access compared to A100 GPU in SNN trainings, and a 20-core deep SNN training and a 5-worker federated learning on FPGAs. Our study develops the first multi-core neuromorphic architecture supporting the direct SNN training, facilitating the neuromorphic computing in edge-learnable applications.

## 1. Introduction

In the wake of escalating demands for intelligent computing, conventional computing architectures, such as GPUs, increasingly grapple with challenges of energy efficiency (Jones, 2018), especially in the energy-constrained edge applications (Alam et al., 2024; Shi et al., 2016; Deepak et al., 2023). Due to the demands for privacy protection and low-latency high-accuracy processing, and limitations of cloud-edge transmission bandwidth, there is a growing necessity for model training at the edge to adapt to dynamic environments (Kaissis et al., 2021; Shao et al., 2024; Li et al., 2021; Babu et al., 2023; Lin et al., 2023). Neuromorphic computing, a paradigm aimed at emulating the structure and functionality of the human brain, offers a promising pathway towards novel, high-energy-efficiency intelligent computing systems (Mehonic & Kenyon 2022; Roy et al., 2019; Zhang et al., 2020; Marković et al., 2020), especially for edge applications (Alam et al., 2024).

Spiking Neural Networks (SNNs), a cornerstone of neuromorphic computing (Maass, 1997), have attracted considerable attention owing to their unique potential for high energy efficiency and fidelity to biological neural processes (Göltz et al., 2021; Li et al., 2024; Stöckl & Maass 2022; Schuman et al., 2022; Roy et al., 2019；Wu et al., 2022). Through the direct training methods based on Back-Propagation (BP) with surrogate gradient learning (Wu et al., 2018; Neftci et al., 2019; Nunes et al., 2022; Dampfhoffer et al., 2023; Zhou et al., 2024), there have been rapid advancements in the performance of deep SNN models in recent years, bringing them close to the performance level of conventional Deep Neural Networks (DNNs) (Fang et al., 2021; Hu et al., 2024; Su et al., 2024; Yao et al., 2023; Yao et al., 2024b; Zhou et al., 2024; Zhang et al., 2024), indicating substantial potentials in intelligent edge applications such as sensory processing, robotics, UAV, embodied intelligence, and federated learning (Bartolozzi et al., 2022; Imam &Cleland 2020; Krauhausen et al., 2021; Salt et al., 2020; Paredes-Vallés et al., 2024; Putra et al., 2024; Sandamirskaya et al., 2022; Venkatesha et al., 2021; Yang et al., 2022; Yao et al., 2024a; Yu et al., 2023).

Neuromorphic processors supporting SNN computation have been developed quickly, such as TrueNorth (Merolla et al., 2014), Loihi (Davies et al., 2018, 2021), Tianjic (Pei et al., 2019), BrainScale (Pehle et al., 2022; Cramer et al., 2022), Neurogrid (Benjamin et al., 2014), DYNAPs (Moradi et al., 2017), Darwin (Ma et al., 2017, 2024). Recent studies have also proposed SNN training architectures H2learn and SATA (Yin et al., 2023; Liang et al., 2022; Bhattacharjee, 2024). Despite these progresses, high-efficiency computation for SNN training still faces three fundamental challenges. Firstly, the architectures of these neuromorphic processors are primary designed for brain simulation, which supports feedforward computations and local learning methods (Merolla et al., 2014；Davies et al., 2021; Pei et al., 2019; Cramer et al., 2022; Ma et al., 2024), such as Hebbian learning and Spike-Timing-Dependent Plasticity (STDP), but none of them support the direct training of deep SNNs, such as convolutional SNNs, based on backpropagation. Currently, SNN training heavily relies on GPUs. Secondly, the recently proposed SNN training architectures (Yin et al., 2022; Liang et al., 2022) are still single-core architectures, lacking the inter-core design and hardware implementation. There is still a lack of multi-core architectures to support efficient model deployment, pipelines, and parallel computation, to meet the demands of deep SNN training. Thirdly, there is a growing demand to reduce the energy consumption of Dynamic Random-Access Memory or High Bandwidth Memory (DRAM/HBM) access during SNN training. With increasing parameters of SNN models, extensive access to DRAM/HBM leads to high power consumption and long latency. This challenge is particularly significant for SNNs, as they involve multiple time steps and more complex data dependencies compared with ANNs, resulting in high storage requirements during their training.

To address these issues, we propose, to the best of our knowledge, the first multi-core, near-memory neuromorphic architecture supporting SNN training based on BP. In this architecture, each computing core contains Feedforward-Propagation (FP), Back-Propagation (BP), and Weight-Gradient (WG) computing engines, achieving highly efficient parallel computing at both the engine and core levels. Through joint design of the computing core, data flow, and network-on-chip (NOC), our multi-core neuromorphic architecture achieves a high level of data reuse during SNN training, reducing DRAM access by ~55 to ~85% compared with A100 GPU. By fully leveraging the sparsity of spikes, derivatives of firing function, and membrane potential gradients, energy consumption is reduced by ~45 to ~60%. By implementing this multi-core training architecture on FPGAs, we achieve 20-core deep SNN training and 5-worker federated learning. Our study extends the common multi-core neuromorphic computation from inference and local learning to global SNN training based on BP.

## 2. Results

Figure 1a highlights the advantages, challenges, and our proposed solutions for neuromorphic computing. The training process of SNN (Figures 1b and 1c) with the Leaky Integrate-and-Fire (LIF) neuron model includes FP, BP, WG stages (Wu et al., 2018; Wu et al., 2019; Gu et al., 2019). The SNN parameters $N$, $T$, $H$, $W$, $C$, $R$, $S$, $M$, $E$, $F$ are described in Suppl. Table 2, and Figure 1d shows examples of membrane potential updating and spike generation in LIF neurons during FP.

### 2.1 Instruction Set
We designed an instruction set to support SNN training (Suppl. Table 1). Convolution instructions include FP, BP, WG convolutions, supporting different kernel sizes, strides, interleaves, and padding configurations (equations (2), (7), and (8) in Methods). Soma/Grad instructions support the LIF model, including Soma operations during FP (equations (1) and (3) in Methods) and Grad operations during BP (equation (4) and (6) in Methods). BN instructions support batch normalization during the FP and BP. Vector instructions support cross-core accumulation.

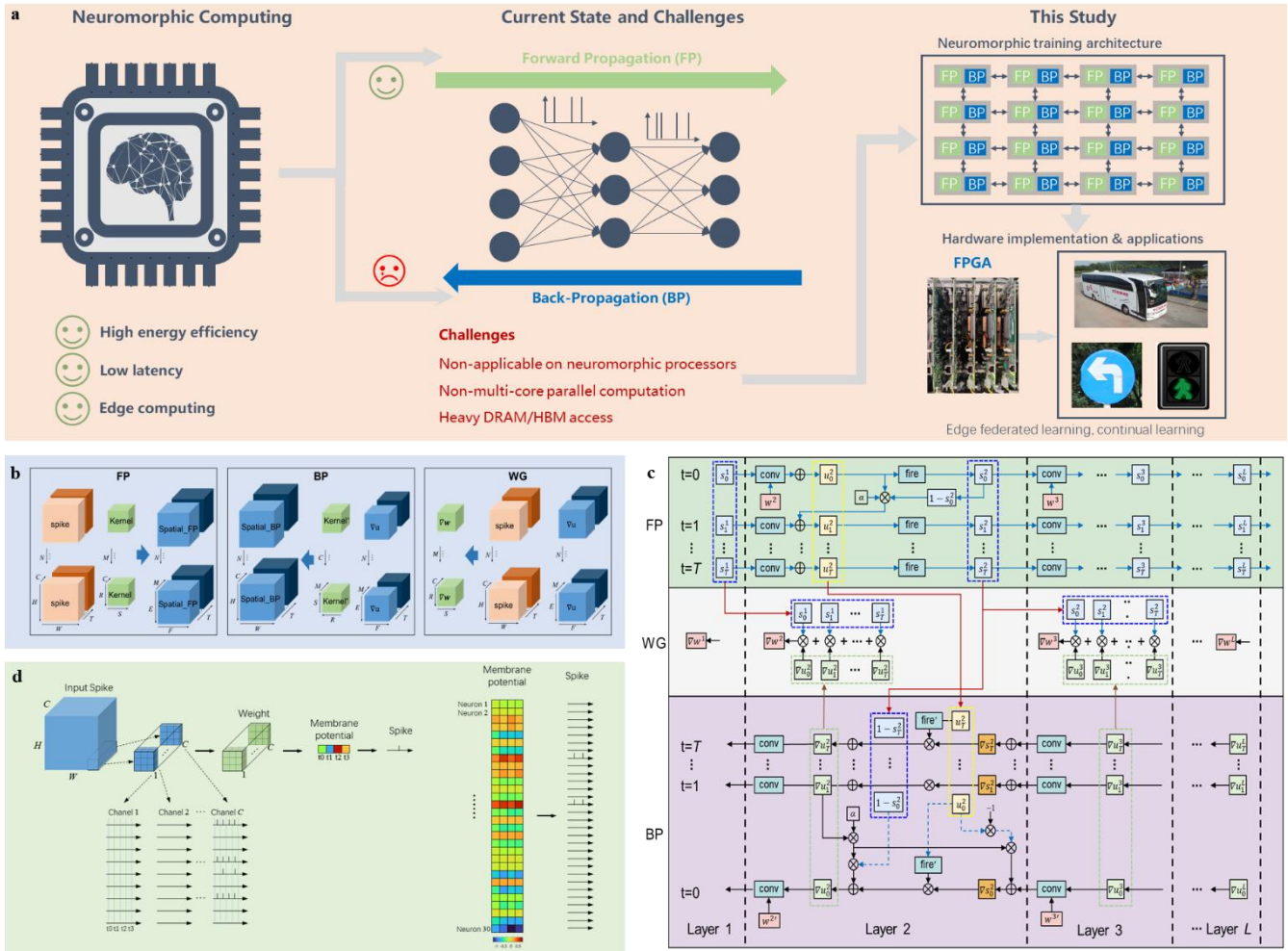

**Figure 1. Neuromorphic computing and SNN.** (a) An overview of neuromorphic computing, highlighting its advantages, challenges, and our proposed solutions. (b) Convolution computations in Feedforward-Propagation (FP), Back-Propagation (BP) and Weight Gradient (WG) processes in SNN training. (c) Computation and data transmission in the FP, BP, and WG processes. Full illustration is shown for layer 2, and detailed computations of Soma and Grad are shown for time step 0. (d) Membrane potential updating and spike generation in LIF neurons: An example LIF neuron in a convolution layer of SNN (left) and 30 LIF neurons in a convolution layer (right).

## 2.2 Multi-Core Near-memory Neuromorphic Architecture for SNN Training

We proposed a multi-core near-memory neuromorphic architecture that supports FP, BP, and WG computations to achieve efficient deep SNN training. The register-transistor level (RTL) design was completed and verified in both software and FPGA hardware. Figure 2a shows the design of a single computing core within the multi-core architecture, including an FP Sub-Core, a BP Sub-Core and a Router. DRAM is shared by all computing cores in the architecture.

The FP computing engine in the FP Sub-Core implements the forward inference of SNN, including FP Array, Soma, Fire, Pooling, and related SRAM (Figure 2a). The FP Array consists of a 16×16 selector and adder array with FP16 precision. Its input includes spike signals $s$ and the weights $w$. Since the spike signals are either 0 or 1, the FP engine performs the conventional multiply-add convolution by selector-gated addition operations, that is, only adders in the column of FP array with input spike value of 1 perform addition on weight values. The FP Array performs accumulation of *w* across the *C, R, S* and *T* dimensions to generate 16 *M* dimension convolution results *Conv_FPs*. The spike signal *s* is represented by a single bit to save storage space. The Soma module (Figure 2b) updates the neuron's membrane potential based on its previous membrane potential, leakage factor *α*, spike firing state, and the *Conv_FP* according to equation (2) in the Methods. The membrane potential is compared with a threshold to generate the output spike signal in its Fire sub-module. An FP engine includes 16 Soma modules, corresponding to 16 *M* dimension channels. The Pooling module supports max pooling operations on the output spike signals.

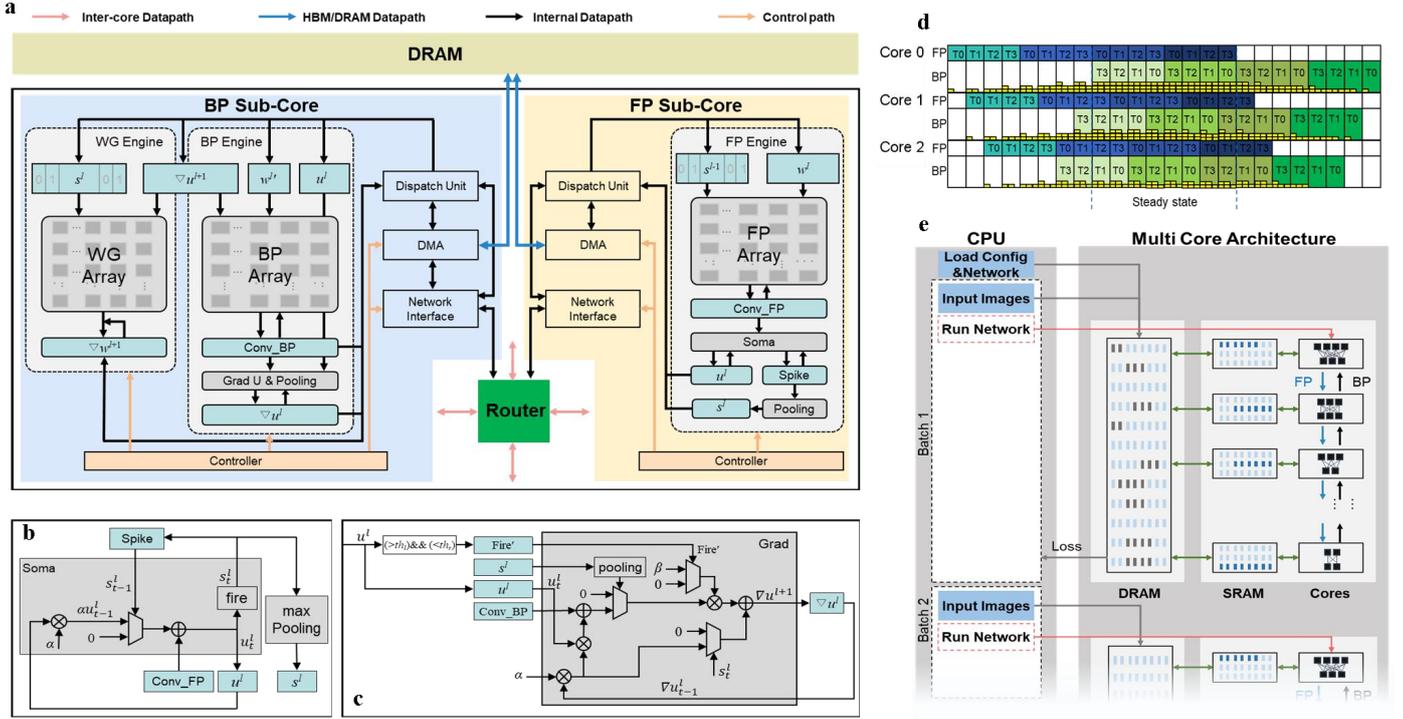

**Figure 2. The Multi-core Neuromorphic Architecture for SNN Training.** (a) Architecture of single computing core with an FP sub-core and a BP sub-core. (b) Soma module in FP engine. (c) Grad module in BP engine. (d) Parallel forward and backward computation pipeline on the multi-core architecture. (e) Training procedure on the multi-core neuromorphic architecture.

The BP Sub-Core includes a BP Engine and a WG Engine (Figure 2a), which computes membrane potential gradients $\nabla u$ and weight gradients $\nabla w$ during backpropagation in parallel. The two engines share the input of $\nabla u$ backpropagated from the next layer. The BP Engine contains BP Array, Grad, Pooling, and related SRAM. The BP Array is a 16×16 MAC array with FP16 precision, calculates products of $\nabla u$ and kernel-rotated weight $w'$ and accumulates these products across the *M, S, R,* and *T* dimensions to generate 16 *C* dimension convolution results *Conv_BPs*. The Grad module (Figure 2c) calculates the spike gradients $\nabla s$, the surrogate gradient $fire'$, and $\nabla u$ according to equation (6), (5), (4) in the Methods, respectively. A BP engine includes 16 Grad modules, corresponding to 16 *C* dimension membrane potential gradients $\nabla u$. The WG engine contains a WG Array and related SRAM. The WG Array consists of a 16×16 adder array and selectors with FP16 precision, performing accumulation of $\nabla u$ across the *T* and *N* dimensions to generate weight gradients $\nabla w$. Its input includes output spike signals *s* obtained during forward computation and $\nabla u$. Only the adders in the column of WG array with input spike value of 1 perform addition on $\nabla u$.

Modules for data transmission in the architecture include the Dispatch Unit (DU), DMA, and Network-On-Chip (NOC) consisting of Router and Network Interface (NI). The DU supports data transmission between on-chip SRAMs of different sub-cores/ cores and data transmission between on-chip SRAM and DRAM in collaboration with NI and DMA. A Router (Suppl. Figure 1a) has 6 ports, 2 of which are interconnected with the FP Sub-Core and BP Sub-Core within the same core, and 4 are interconnected with 4 adjacent computing cores. To support SNN training, the multi-core architecture includes three primary on-chip data transmission channels: 1) FP-FP transmission transferring the spike output $s$ of a layer to the FP sub-core processing the next layer; 2) BP-BP transmission transferring the output $\nabla u$ of a layer to the BP sub-core processing the previous layer; 3) FP-BP transmission transferring the spike $s$ and membrane potential $u$ obtained in forward computation to BP sub-cores. These data transmission channels support parallel forward and backward computing across multiple cores (Figure 2d).

We implemented clock domain isolation to decouple the working frequency of Routers and computing cores, with the computing core operating at 500 MHz and the Routers operating at a higher frequency of 667 MHz in the RTL simulation to increase data transmission bandwidth. The Controller unit in each sub-core monitors its working status and coordinates the orderly operation of modules in the sub-core. The multi-core neuromorphic architecture was designed with 32 cores containing 64.78MB SRAM, interconnected via an 8×4 2D-mesh NOC network (Suppl. Figure 1b), achieving 16 TFLOPS@ FP16 (Suppl. Table 3).

In contrast to the training procedure on GPUs, where only a portion of a model is deployed at a time, requiring multiple times of instruction transmission from CPU to perform the training on one batch of samples (Suppl. Figure 3), the entire SNN is deployed on our multi-core architecture with different layers allocated on different computing cores, requiring only once of instruction transmission from CPU to perform training on one batch of samples (Figure 2e). Training loss is used for CPU to minitor the training process. Our training procedure facilitates efficient data transfer and reuse across computing cores, reducing DRAM access during training.

## 2.3 Data Flow and Parallel FP-BP-WG Computations

We designed multiple data flows for FP, BP, and WG computations to enhance data reuse and reduce energy consumption of data transmission during training. The FP and BP engines adopt Weight Stationary (WS) data flow and partial sum SRAM reuse (Figures 3a-b). During computation, weights $w$ are retained in the PE arrays, while input $s$ during FP and $\nabla u$ during BP of different spatial positions and time steps are fed into the PE arrays. All weights are read only once from SRAM during the whole convolution computation. The WG engine adopts an Output Stationary (OS) data flow with products of $\nabla u$ and $w'$ retained and accumulated in WG array (Figure 3c). We also designed storage reuse of Conv_FP and Soma in FP engine, Conv_BP and Grad in BP engine, and $\nabla w$ in WG engine (Figures 3a-c).

Figures 3d-f show the high utilization of FP, BP, and WG engines during Spiking-ResNet18, -ResNet50, and -VGG16 training in the software simulations (see Methods), supporting the highly parallel computations of FP, BP, and WG engines within a computing core. In the software simulations, our 32-core architecture achieved 18.20%, 15.88%, 17.47%, 17.14% of A100 GPU performance (frame per second, fps) during Spiking-ResNet18 training (Figure 3g) with batchsize 16, 32, 64, 128, respectively, and 40.11%, 39.48%, 38.46%, 38.08% of A100 during Spiking-ResNet50 training (Figure 3h), and 18.69%, 19.08%, 19.52%, 23.64% of A100 during Spiking-VGG16 training (Figure 3i), although our architecture consisted only 5.13% of A100 computing capability.

The data flow designs enhanced on-chip data reuse and reduced the DRAM occupancy and access. We analyzed HBM/DRAM occupancy and accesses of our 32-core architecture and A100 GPU during Spiking-ResNet18, -ResNet50, and -VGG16 training. Compared with A100, the DRAM occupancy of our architecture was reduced by 41.57%, 50.31%, 55.85%, and 59.00% during Spiking-ResNet18 training of batchsize 16, 32, 64, and 128 (Figure 3j), respectively, by 44.62%, 52.69%, 57.65%, and 60.42% during

Spiking-ResNet50 training (Figure 3k), and by 59.83%, 67.83%, 72.53%, and 75.10% during Spiking-VGG16 training (Figure 3l). DRAM access in our architecture was reduced by 56.65%, 57.54%, 57.71%, and 57.53% during Spiking-ResNet18 training (Figure 3m), by 82.53%, 84.57%, 85.41%, and 85.89% during Spiking-ResNet50 training (Figure 3n), and by 82.93%, 84.41%, 85.28%, and 85.97% during Spiking-VGG16 training (Figure 3o). As the model parameters and batchsize increased, our architecture was more effective in reducing DRAM occupancy and access. The on-chip SRAM size of our 32-core architecture (64.78MB) was smaller than the SRAM size of A100 (88.75MB, NVIDIA A100 Whitebook)

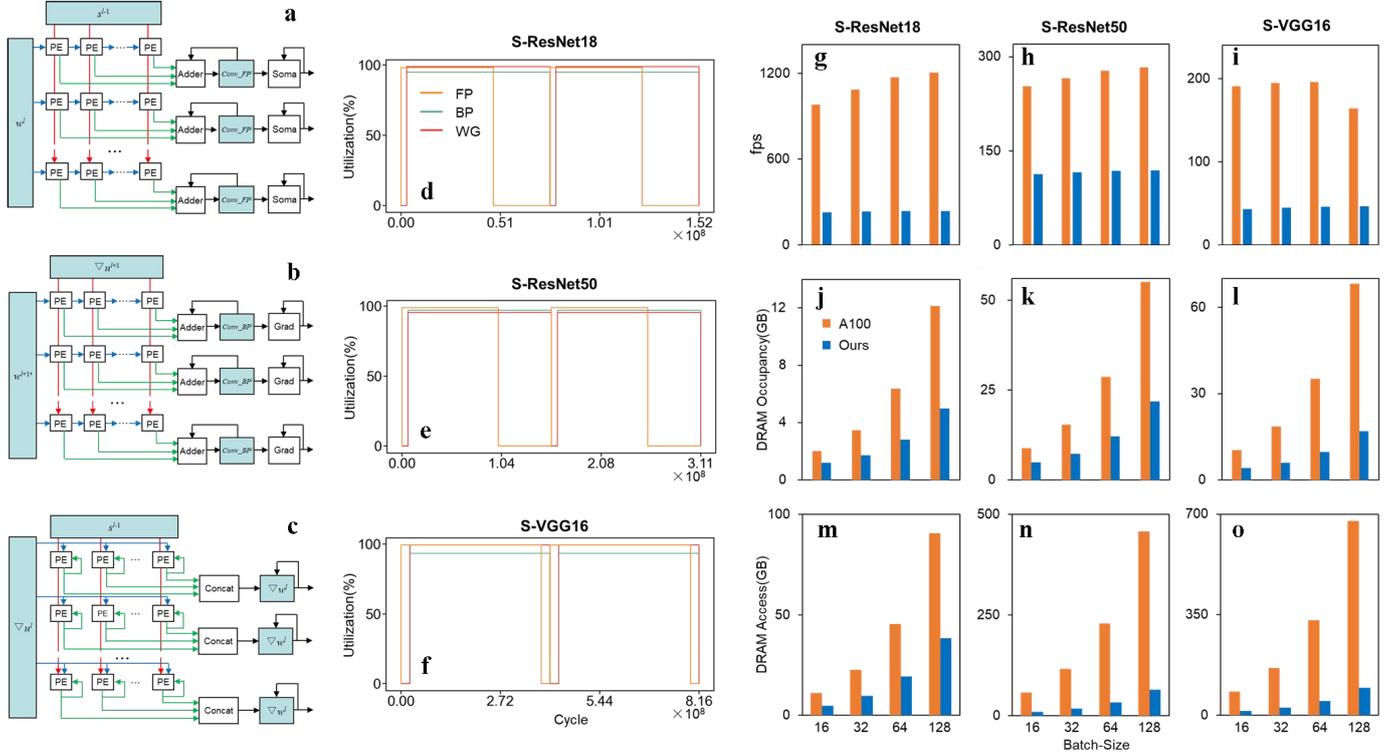

**Figure 3. Dataflow and Parallel Computations in the Multi-core Neuromorphic Architecture.** (a-c) show dataflows of FP, BP, and WG engines, respectively. (d-f) show the utilization of FP, BP, WG engines and computing core during Spiking-ResNet18, -ResNet50, and -VGG16 training, respectively. Utilization curves during two batches of training period are plotted. (g-i) show the comparison of training performance (fps) during the three training, respectively. (j-l) show comparison of DRAM/HBM occupancy between our multi-core architecture and A100 GPU during Spiking-ResNet18, -ResNet50, and -VGG16 training, respectively. (m-o) show the comparison of DRAM/HBM access during the three training, respectively.

## 2.4 Sparse Computation Optimization

We designed sparse control circuits for the three engines. Figure 4a shows the design for a single output channel of the FP engine, which leverages the spike sparsity to control the first-stage adders (gating1) to perform addition only on the weights of the channels with spike $s = 1$. When all 16 channles of $s$ are zero, the entire adder tree computation and partial-sum update are skipped through the gating1 and gating2 signals. Figure 4b shows the design of a single output channel of the BP engine. When $fire'$ is zero, the entire multiplier array and adder tree computation, $\nabla u$ reading, and partial-sum update are skipped through the gating1 signal. The gating2 signal and gating1 signal go through an AND operator to generate the gating signals for 16 multipliers and control signals for selectors. When the $\nabla u$ input of a multiplier is zero, its computation is skipped and its output is selected as zero. In WG engine (Figure 4c), when a channel of s is zero, the corresponding column of 16 accumulators skip accumulation through gating1 signal. When all 16 channels of $s$ are zero, $\nabla u$ reading is also skipped through the gating2 signal. Together, we leverage the sparsity of $s$, $fire'$, and $\nabla u$ signals in our design to reduce redundant computation and SRAM accesses. The RTL power analysis showed that our sparse design could significantly reduce the energy consumption of the computing core during spiking-ResNet18, -ResNet50, and -VGG9 training by 61.09%, 46.81%, and 51.69%, respectively (Figures 4d-f), including 42.53%, 27.01%, 37.11% of reductions in the FP engine, 67.74%, 53.92%, 55.75% of reductions in the BP engine, and 60.16%, 46.24%, 56.52% of reductions in the WG engine.

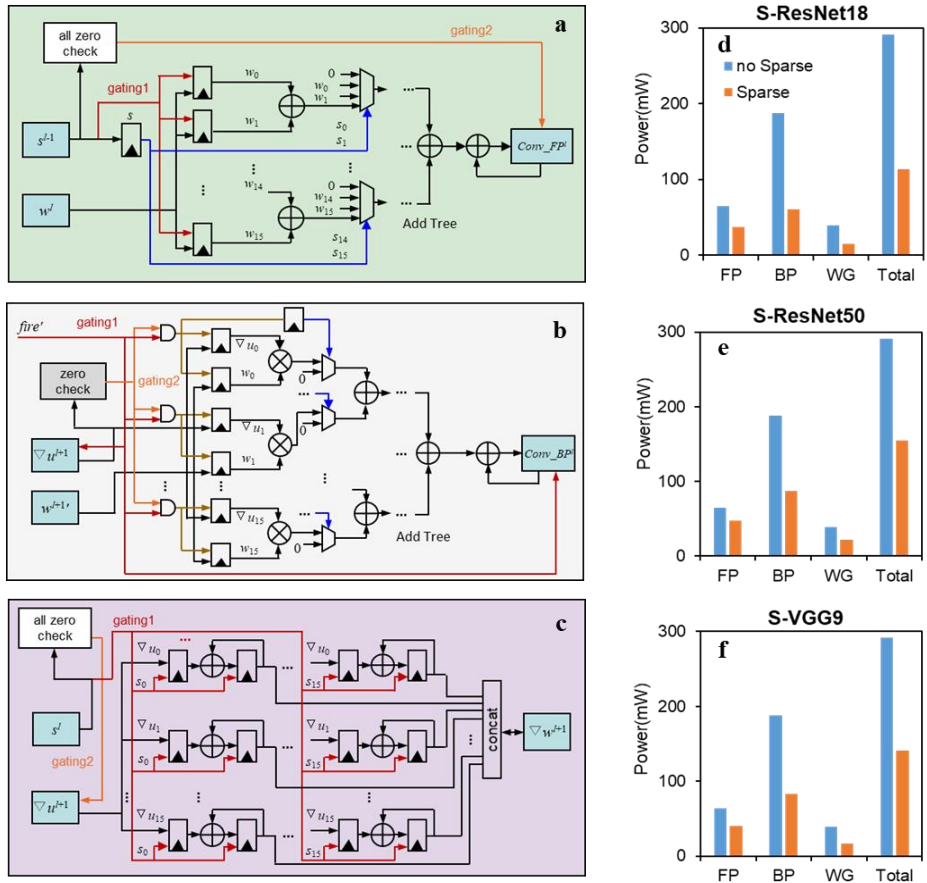

**Figure 4. Sparse Computation Optimization.** (a-c) Sparse computation optimization circuit designs for the FP engine, BP engine, and WG engine, respectively. The red and orange lines serve as gating signals of computation units and SRAM accesses, and blue lines as control signals for Multiplexers. (d-f). Comparisons of the energy consumption before and after sparse optimization for Spiking-ResNet18, -ResNet50, and -VGG9 training, respectively.

**Table 1.** Comparison between our work and representative architectures

|  | Loihi2 | TianjicX | True North | Brain Scale2 | Darwin3 | Graph core | DOJO | GPU A100 | SATA | This Work |
|---|---|---|---|---|---|---|---|---|---|---|
| Process | 7nm | 28nm | 28nm | 28nm | 22nm | 7nm | 7nm | 7nm | 65nm | 28nm |
| Network | SNN | ANN/SNN | SNN | SNN | SNN | ANN | ANN | ANN/SNN | SNN | SNN |
| Frequency (MHz) | 300 | 400 | 1000 | NA | 333 | 1850 | 2000 | 1540 | 400 | 500 |
| On-chip memory (MB) | 24.6 | 22.5 | 260 | NA | NA | 368 | 440 | 88.38 | 4 | 64.78 |
| Area (mm²) | 31 | 64 | 430 | 65 | 358.527 | 832 | 645 | 826 | NA | 218.63 |
| Power (W) | 0.1 | <7.5 | 0.065 | 0.2 | 1.8 | 300 | 400 | 400 | NA | 14.49 |
| Peak Performance (TOPS) | 0.3 | NA | 0.0581 | NA | NA | 250 | 362 | 312 | NA | 16 |
| Precision | INT1-9 | INT8 | INT1 | INT6 | INT1-16 | BF16 | BF16 | BF16 | INT8 | FP16 |
| Energy Efficiency (TOPS/W) | 3 | 3.2 (0.84V) | 0.4 | NA | NA | 0.83 | 0.91 | 0.78 | NA | 1.05 |
| Training Based on BP | no | no | no | no | no | yes | yes | yes | yes | yes |

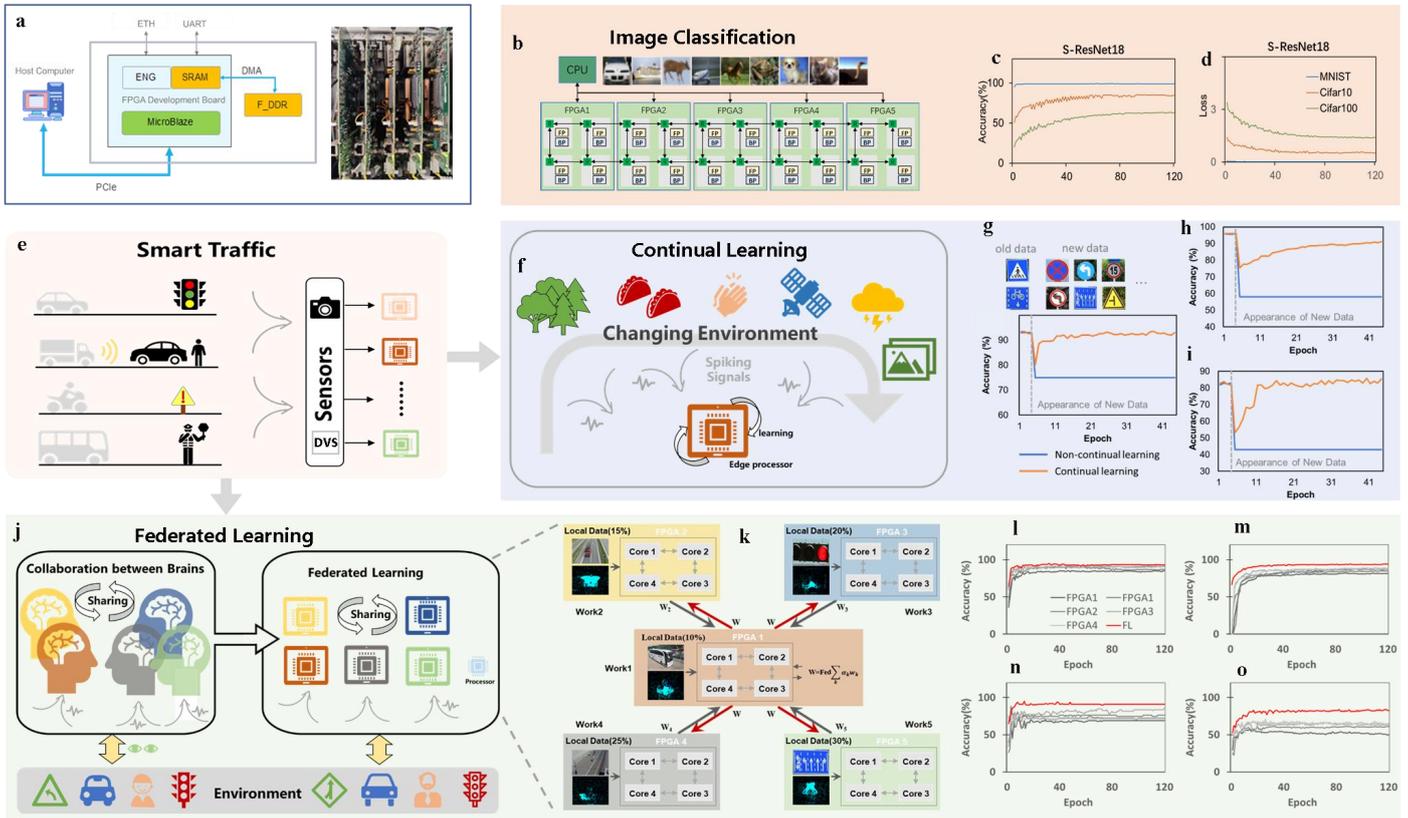

**Figure 5. FPGA Implementation of the Multi-core Neuromorphic Architecture and applications.** (a) FPGA platform. (b) Implementation of a 20-core neuromorphic architecture on FPGAs. (c)-(d) Accuracy and loss curves of SNN training on FPGAs in the image classification tasks. (e) Illustration of smart traffic scenario. (f) Illustration of edge continual learning. Effects of continual learning on the inference accuracies in the Traffic Sign Classification and Recognition dataset (g), Vehicle and Pedestrian Detection Dataset (h), and DVS-Gesture dataset (i) on FPGAs. (j) Illustration of federated learning at the edge. (k) Implementation of the 5-worker federated learning on FPGAs, with each worker implemented on one FPGA and worker1 also used as the server. (l)-(o) Accuracy curves of the five workers and the federated model on the Traffic Sign Classification and Recognition dataset, the Vehicle and Pedestrian Detection dataset, the Pedestrian Augmented Traffic Light Dataset, and the DVS-Gesture dataset.

## 2.5 Area and Power Consumption Analysis

The area and power consumption of a single computing core is estimated in RTL simulation. The BP, FP, and WG Engine together accounted for 78% of the total area and 63% of the total power consumption, while the rest modules including Router, NI, DMA, DU, and Controller, together accounted for 22% of the total area and 37% of the total power consumption (Suppl. Figure 5a). Grouped into computation, storage, and transmission, the on-chip storage occupied the largest area of the computing core (67%), while the power consumption of the three groups is comparable (30~38%; Suppl. Figure 5b).

Table 1 shows the performance comparison between our 32-core near-memory neuromorphic architecture and representative architectures such as neuromorphic chips and GPUs. Our architecture achieved an energy efficiency of 1.05 TFLOPS/W @ FP16 @ 28nm, which is comparable to the A100, DOJO, and Graphcore with 7nm technology, demonstrating the high energy efficiency of our architecture.

## 2.6 FPGA Implementation and Applications

The multi-core near-memory neuromorphic architecture was deployed on an FPGA platform (Figure 5a) to perform SNN training through our software toolchain (Suppl. Figure 4). The detailed hardware utilization is listed in Suppl. Table 4. Figure 5b shows that a 20-core neuromorphic architecture was deployed on five FPGA cards interconnected with 2×10 2D Mesh topology. Training of spiking-ResNet18 deployed on the 20-core architecture converged normally in image classificaitons tasks on MNIST, Cifar10, and Cifar100 datasets (Figures 5c-d), with accuracies 99.23%, 85.5%, and 63.89%, respectively. The accuracy loss compared to GPU was 0.05%, 0.49%, and 2.41%. We then implemented continual learning (Figure 5f) and federated learning (Figure 5j) applications in the FPGAs for a smart traffic seneiro (Figure 5e). The continual learning improved inference accuracy on mixed new and old data by 17.97% for the Traffic Sign Classification and Recognition dataset (Figure 5g), 33.21% for the Vehicle and Pedestrian Detection dataset (Figure 5h), and 42.41% for the DVS-Gesture dataset (Figure 5i). In the federated learning application, five FPGAs served as five workers (4 cores per worker) (Figure 5k), and the worker1 is also used as the server. Figures 5l-o show that the federated model significantly outperformed worker models, achieving accuracies of 94.75%, 94.09%, 94.85%, and 84.03% on the Traffic Sign Classification and Recognition dataset, the Vehicle and Pedestrian Detection dataset, the Pedestrian Augmented Traffic Light dataset, and the DVS-Gesture dataset, respectively.

## 3. Discussion

To the best of our knowledge, we have developed the first multi-core neuromorphic architecture supporting deep SNN training based on BP. Each computing core in our architecture contains the FP, BP, and WG engines that support high parallelism of forward and backward computations during SNN training. Our architecture adopts WS and OS data flows, on-chip SRAM data reuse, and bit storage of spike signals, achieving a 70~90% reduction of DRAM/HBM access compared to A100 GPU during SNN training. It implements sparse computation optimizations for all three engines to fully leverage the sparsity of signals in SNN training, reducing energy consumption by ~45% to 60% through skipping redundant computations and data movements. It achieves a high energy efficiency of 1.05 TFLOPS/W @ 28nm, which is comparable to that of A100, DOJO, and Graphcore with 7nm technology. The FPGA implementation of our multi-core neuromorphic architecture in continual learning and federated learning scenarios demonstrates its potential for edge-learnable applications.

Existing neuromorphic processors, such as TrueNorth (Merolla et al., 2014), Loihi (Davies et al., 2018, 2021), Tianjic (Pei et al., 2019), BrainScale (Pehle et al., 2022; Cramer et al., 2022), Neurogrid (Benjamin et al., 2014), DYNAPs (Moradi et al., 2017), and Darwin (Ma et al., 2017, 2024), have successfully implemented SNN inference and local learning. However, training these networks based on BP remains a significant challenge for these processors (Bellec et al., 2020; Journ'e et al., 2022). Most neuromorphic computing approaches rely exclusively on off-chip learning (Rueckauer et al., 2022; Meng et al., 2022; Yin et al., 2023) or implement a hybrid training loop, where on-chip inference is combined with off-chip BP performed on conventional computers (Cramer et al., 2022; Friedmann et al., 2016). To perform SNN training on these neuromorphic processors, people also develop alternatives or approximations to BP (Bellec et al., 2020; Lee et al.,2020; Payvand et al., 2020; Frenkel et al., 2020; Shrestha et al., 2021), as well as methods combining local learning rules and global signals such as last layer errors (Kreiser et al., 2017; Journ'e et al., 2022; Xiao et al., 2024), while falls short in training accuracy (Meng et al., 2023). Recently, Renner et al. (2024) proposes a method combining a gating mechanism with Hebbian learning to enable BP for a single hidden-layer fully connected network on Loihi. However, its applicability to popular SNNs, such as convolutional SNNs, remains a question. Additionally, it consumes over 200 times the energy of inference, raising concerns about its energy efficiency. In the recently proposed SNN training architecture SATA, its FP, BP and WG computations share the same PE array, excluding the parallel computation of the three processes. H2Learning employs a high-efficiency look-up tables approach for SNN training. However, it cannot complete the SNN training process alone, which depends on other computing chips/platforms to calculate the items in the look-up tables. In addition, both SATA and H2Learning lack multi-core designs and hardware implementation.

## 4. Methods
### 4.1 SNN Training Algorithm
The SNN adopts the LIF neuron model, and the training process is divided into the following stages (Y. Wu et al., 2018; Y. Wu et al., 2019; Gu et al., 2019).

**Forward Propagation (FP) Stage:** The membrane potential and output spike signal during the SNN's feedforward inference process are calculated. $u_t^l[i]$ and $s_t^l[i]$ represent the membrane potential and spike signal of neuron $i$ in layer $l$ and time step $t$, respectively. $u_t^l[i]$ is determined by the state of the neuron at time step $t$-1 and the membrane potential change caused by its synaptic input, calculated as:

$$u_t^l[i] = \alpha u_{t-1}^l[i](1 - s_{t-1}^l[i]) + Conv\_FP_t^l[i] \quad (1)$$

where $Conv\_FP_t^l[i]$ is the convolution of the spike output signal of layer $l$-1 and the weight value of the current layer $l$:

$$Conv\_FP_t^l[i] = \sum_j s_t^{l-1}[i] w^l[j,i] \quad (2)$$

$\alpha$ is the leakage factor of the membrane potential. The membrane potential of the neuron is then compared with the threshold voltage to obtain its spike output signal $s_t^l[i]$ through the spike activation function $fire(u_t^l[i])$:

$$s_t^l[i] = fire(u_t^l[i]) = \begin{cases} 1, & u_t^l[i] \geq th_f \\ 0, & else \end{cases} \quad (3)$$

When the membrane potential exceeds the threshold voltage, the neuron emits a spike and resets the membrane potential to zero.

**Back Propagation (BP) Stage:** The membrane potential gradient $\nabla u_t^l[i]$ and spike gradient $\nabla s_t^l[i]$ during the Back-Propagation process are calculated as follows:

$$\nabla u_t^l[i] = \alpha \nabla u_{t+1}^l(1 - s_t^l[i]) + \nabla s_t^l[i] fire'(u_t^l[i]) \quad (4)$$

Where the $fire'(u_t^l[i])$ is

$$fire'(u_t^l[i]) = \begin{cases} 1, & th_l \leq u_t^l[i] \leq th_r \\ 0, & else \end{cases} \quad (5)$$

The spike gradient $\nabla s_t^l[i]$ is calculated as:

$$\nabla s_t^l[i] = -\alpha \nabla u_{t+1}^l[i] u_t^l[i] + Conv\_BP_t^l[i] \quad (6)$$

where $Conv\_BP_t^l[i]$ is the convolution of the membrane potential gradient and the weight value of layer $l$+1:

$$Conv\_BP_t^l[i] = \sum_j \nabla u_t^{l+1}[j] w^{l+1}[i,j] \quad (7)$$

**WG Stage:** The weight gradient $\nabla w^l[i,j]$ is calculated based on the membrane potential

gradient of current layer and the spike output signal of layer *l*-1 obtained during the forward computation:

$$\nabla w^l[i,j] = \sum_t \nabla u_t^l[j] s_t^{l-1}[i] \quad (8)$$

Finally, the weights are updated using $\nabla w^l[i,j]$.

## 4.2 Multi-core Neuromorphic Architecture Design for SNN Training

### A. Multi-core Neuromorphic Architecture Design

We designed a multi-core near-memory neuromorphic architecture that supports FP, BP, and WG computations for deep SNN training. The single computing core within the multi-core architecture includes FP Sub-Core, BP Sub-Core, and Router.

The FP Array in the FP Sub-Core consists of a 16×16 selector and adder array with FP16 precision. The 16 *C* dimension input spike signals *s* are broadcasted to 16 columns of the FP Array, respectively. The FP Array performs parallel computations of 16 output channels along *M* dimension vertically and accumulation across 16 different input channels along the *C* dimension horizontally, and once outputs 16 channel partial-sums of weights *w* accumulated across the 16 *C* dimensions, and further accumulates these weight partial-sums across the *R*, *S*, and *T* dimensions to generate 16 *M* dimension convolution results *Conv_FPs* by refreshing the input spike signals *s* and weights *w*. Through the spike signals' control of selectors, only adders in the column of FP array with input spike value of 1 perform addition on weight values. For the Soma module, if there was no spike firing in the previous time step, the membrane potential at the previous time step is first multiplied by its leakage factor *α* and added to *Conv_FP* to get the membrane potential at current time step. Otherwise, the membrane potential at current time step is set to the *Conv_FP*.

The BP Array in the BP Sub-Core is a 16×16 MAC array with FP16 precision. Its input includes 16 membrane potential gradients $\nabla u$ backpropagated from next layer and 256 kernel-rotated weights *w'*. The membrane potential gradients of 16 output channels along the *M* dimension are broadcasted to the 16 columns of the BP Array, respectively. The BP Array performs parallel computations on 16 the input channels along the *C* dimension vertically and accumulation across 16 *M* dimension output channels horizontally, and once outputs 16 *C* dimension channel partial-sums of products between $\nabla u$ and *w'* accumulated across the 16 *M* dimensions. By refreshing input membrane potential gradients $\nabla u$ and weights *w'* multiple times, the BP Array accumulates the partial-sums across the *M*, *S*, *R*, and *T* dimensions to obtain the convolution result *Conv_BP*.

The WG Array in the WG engine consists of a 16×16 adder array and selectors with FP16 precision. 16 *C* dimension channels of input spikes *s* are broadcasted to 16 columns of the WG Array, respectively. 16 *M* dimension channels of membrane potential gradients $\nabla u$ are broadcasted to 16 rows of the WG Array, respectively. Each column of the WG Array shares a selector controlled by the input spike signal. Thus, only the adders in the column of WG array with input spike value of 1 perform addition on membrane potential gradients. The WG Array first computes the weight gradients for one timestep, and accumulates the gradients of different time steps to obtain $\nabla w$ according equation (8).

Modules for data transmission in the architecture includes the Dispatch Unit (DU), DMA, and Network-On-Chip (NOC) consisting of Router and Network Interface (NI). The DU consists of a cache, read & write arbiters, bit-width converters, and transceivers, supporting data transmission between on-chip SRAMs of different Sub-Cores within the same core or across cores and data transmission between on-chip SRAM and DRAM in collaboration with NI and DMA, respectively. The NI is the interface module of the NOC network, including data caches, connection configuration tables, transceivers etc. Data for transmission are encapsulated into NOC packets in the NI with routing information. NOC packets are routed to NI of the target core through the Router, and the NI parses the NOC packets and sends the data to the

DU or DMA. Each computing core contains a Router, and each Router has 6 input and output ports, 2 of which are interconnected with the FP Sub-Core and BP Sub-Core within the core, and 4 are interconnected with 4 adjacent computing cores. The Controller in each sub-core monitored the working status of the sub-core's modules, and coordinated the orderly operation of all modules (Suppl. Figure 2) based on the configuration information generated by model compilation. The multi-core near-memory architecture was designed to contain 32 cores interconnected via a 4×8 2D-mesh NOC network.

### B. Architecture Evaluation Based on Software Simulation

To quantitatively evaluate the performance of the multi-core near-memory computing architecture during SNN training, we developed a software tool S-ZigZag based on ZigZag, a performance evaluation tool for computing architecture during ANN inference [Mei et al., 2021]. S-ZigZag supports the evaluation of energy consumption, throughput, and latency for different data flows during SNN training tasks under given computing architecture. All software tools are implemented under Python (version 3.8.17).

SNN BP convolution and ANN convolution both use floating-point convolution. Therefore, in S-ZigZag, we adopted ZigZag's convolution computation simulation module to simulate the BP convolution of one timestep in SNN, with multiple simulations for multiple time steps of BP convolution. For FP and WG convolution computations of SNN, we developed new simulation modules to simulate convolution computation between single-bit spike and floating-point weight data using selector and adder arrays. We developed modules for Soma and Grad computation simulation. Soma simulation includes selection, addition, multiplication, and comparison operations, while Grad simulation includes selection, addition, and multiplication operations. We also developed an NOC module to simulate the data transmission volume and energy consumption between computing cores.

In S-ZigZag, we adopted *Json* format for the description files. The computing architecture description file includes the type, count, and energy consumption information of computation parts in the three engines; capacity, bandwidth, datawidth, bank number, and read/write energy consumption information for DRAM, SRAM, and REG three-levels of storage. The model description file includes dimension information for all layers of SNN models. The data flow description file described the temporal and spatial mapping on the computation array through the for-loop form.

$P$ represents the number of cycles to train one batch of data, as follows:

$$P = P_{FP}^{1st} + max(P_{BP}, P_{WG}) \qquad (9)$$

where $P_{FP}^{1st}$ represents total cycle number of FP processing the first data of each batch, and $P_{BP}$ and $P_{WG}$ represent total cycle number of BP and WG processing a batch data.

The utilization rate of the computing core is:

$$Util = \frac{\sum_{i=1}^{P}(Op_{FP\_used}^i + Op_{BP\_used}^i + Op_{WG\_used}^i)}{P \times Op_{total}} \qquad (10)$$

where $Operator_{FP\_used}^i$, $Operator_{BP\_used}^i$, and $Operator_{WG\_used}^i$ represent the number of active multiplication and addition components in FP, BP, and WG during cycle *i*, respectively. $Operator_{total}$ represents the total number of computing components of the three engines.

**Comparison with GPU**: We compared our 32-core near-memory computing architecture with the NVIDIA A100 80GB GPU in DRAM/HBM memory usage and access during SNN training. Our 32-core computing architecture has a total of 64.78MB on-chip SRAM storage, while the A100 has 88.38MB. We compared the peak DRAM usage and total accesses of one batch training with the ImageNet2012 dataset. Batch sizes include 16, 32, 64, and 128, and trained models are spiking_ResNet18, _ResNet50, and _VGG16. On the A100 GPU, we implemented the BPTT-based SNN model using Pytorch (Zheng et al., 2021) and obtained the peak HBM usage during SNN training through the API (https://pytorch.org/docs/2.0/generated/torch.cuda.max_memory_allocated.html). We used NVIDIA's Nsight Compute tool (https://developer.nvidia.com/nsight-compute) to collect volumes of

HBM access through the ram__bytes_read.sum and dram_bytes_write.sum, and summed the statistics of all CUDA kernel functions to get the total access volume for training one batch. SNN models were split into 32 parts and deployed them on 32 cores in a balanced manner to balance the total delay of each core during model training. We simulated training one batch, considering the on-chip SRAM size and the data volume required for intra-core computation. Using the S-ZigZag tool, we analyzed the storage locations and access counts of all variables during FP, BP, and WG processes to obtain the total data access volume and peak DRAM usage for training one batch.

### C. RTL Coding and Functional Verification

We developed Verilog code in the *gvim* editor to implement the multi-core near-memory computing architecture. The computing core was coded separately according to its modules, then these modules were packed together. Data flow and sparse optimization were also coded in the relevant modules. The total codes for a single computing core were about 30,000 lines. The codes were compiled and synthesized using Design Compiler (DC, Synopsys Inc), and formed an independent IP. To implement the 32-core architecture, 32 copies of the IP were interconnected via an 8×4 2D-mesh NOC network.

We used the Verilog Compiled Simulator (VCS, Synopsys Inc) to perform multi-platform verification at single module, single computing core, and multi-core architecture levels, verifying the correctness of computation, data transmission, control functions, and software configuration. At the single module level, we designed test cases specific for each module including convolution operations with different kernel sizes and strides, and Soma calculations, partial sum accumulation for the FP/BP/WG engine modules, data transmission for DU, Router, NI, and DMA, and status detection and state jump for the control module. At the single computing core level, we designed test cases including multi-timestep forward and backward computations of a single SNN layer for the FP sub-core and BP sub-core, $s/u$ transmission across sub-cores, and multi-path data transmission for NOC and DMA. At the multi-core architecture level, SNN models were deployed layer by layer across different computing cores on the RTL simulation platform. The forward and backward RTL simulation outputs for all layers were compared with the SNN training simulation results (Golden values) based on the C language.

The effectiveness of the Weight Stationary (WS) data flow in FP and BP engines was verified by analyzing the SRAM accesses and PE retaining of weights during forward or backward computations. The weight value in the FP and BP engines was read from SRAM only once during convolution computation and retained in the PE. The effectiveness of the output stationary data flow of the WG engine was verified by analyzing the iterative accumulation values of products of $\nabla u$ and $s$ in each PE. The output results of all PEs remained in the PE, verifying the output stationary data flow of the WG engine. By counting the SRAM accesses of *Conv_FP* in FP engine, *Conv_BP* and $\nabla u$ in BP engine, and $\nabla u$ in WG engine, we verified the effectiveness of the sparse design of the three engines.

### D. Area and power consumption analysis

We used DC and VCS (Synopsys Inc.) to evaluate the area and power consumption of our multi-core neuromorphic architecture. The computing core was divided into modules for analysis. We generated gate-level netlists and area reports based on the RTL code with TSMC 28nm process.

Power analysis was based on the gate-level netlists using VCS. Test cases were constructed based on data sparsity during the actual training process of Spiking-ResNet18, -ResNet50, and -VGG9 models. The spike sparsity for Spiking-ResNet18, -ResNet50, and -VGG9 was 90%, 80%, and 88%, respectively. The *fire'* sparsity was 68%, 54%, and 56%, and $\nabla u$ sparsity was 4%, 1%, and 62%, respectively. VCS simulation outputs waveform files in FSDB format, which were converted to toggle rate files of SAIF format. Power reports were generated based on gate-level netlists and SAIF files using DC.

We constructed test cases with 0% sparsity to obtain power consumption without any sparse optimization ($P_0$). To eliminate the influence of power consumption unrelated to sparsity, we also constructed test cases with 100% sparsity and obtained its power consumption ($P_{100}$). If the power consumption of SNN test case is $P_{SNN}$, the percent of power reduction by sparse optimization for the SNN case was defined as: ($P_0$ - $P_{SNN}$) / ($P_0$ - $P_{100}$) × 100%.

### 4.3 SNN Compilation and Deployment on the Multi-core Neuromorphic Architecture

We developed a lightweight deep learning software framework using *Eclipse* by the C language, in which SNN models are represented through computation graphs expressed by a basic operator library. The computation graphs of a SNN model were split into feedforward and backpropagation sub-graphs according to convolutional layers. One sub-graph was deployed on one sub-core, and adjacent layers were deployed on adjacent cores in the multi-core architecture. The sub-graphs, after the basic operator fusion, were first converted into the custom-designed intermediate representation (IR) described in the C language. Then, the IR was compiled to generate device-side files for each sub-graph, including task description of the sub-core to compute a sub-graph. Driver running on the Controller unit parsed device-side files and generated detailed implementation information for each sub-core, such as instruction's opcodes and operands, instruction sequence, input and output data orchestration in SRAM and HBM, and wrote them into Registers of the sub-core. In our multi-core architecture, detailed implementation information for all layers in sub-cores were generated before the training started, therefore, it required instruction information transmission only once from host to perform training on one batch of samples.

### 4.4 FPGA Implementation

The multi-core near-memory neuromorphic architecture was implemented on a platform with five FPGA boards (VCU128, Xilinx) and a CPU server (ThinkSystem SR670, Lenovo). FPGAs were connected with the CPU via the PCIe 3.0 interface. All BP, FP, WG convolution computations during SNN training were performed on FPGAs. Each FPGA board deployed four computing cores. Each FP16 adder or FP16 multiplier consisted of 1 DSP. Each computing core consumed 1183 DSPs, 216K LUTs, 209K FFs, 244 BRAMs, and 104 URAMs. The single core Verilog code was encapsulated into custom IP by the FPGA development tool Vivado (Xilinx). MicroBlaze of FPGA was used as the Controller unit of each sub-core. FPGA boards were interconnected through Routers and Ethernet interface. We developed a Transport Layer Protocol for the Ethernet interface, and RTL codes for clock and reset management circuits to achieve correct synchronization and reset of each computing core. NOC adopted a streaming protocol and multi-port transmission scheduling based on flit segmentation. The computing core operated at 150 MHz and Router operated at 200 MHz achieving 50×6 Gbps bandwidth with its six ports. All RTL codes and IPs were synthesized first, then, after place and route, bitstream files were generated in Vivado and downloaded to Configuration RAM (CRAM) on FPGA for testing and verification using JTAG tools. Finaly, the bitstream files were burned into the Flash memory in each FPGA board.

We developed a driver for initializing all sub-cores in FPGA, parsing the device-side files of each computing sub-core, and writing the corresponding instruction information into registers of the sub-core. The driver was running on the Controller unit (MicroBlaze) of each sub-core. Model weights and dataset were first loaded into the HBM from the host, then transferred to SRAM for computations by dynamically calling the DMA instructions during training.

For the smart traffic scenario, open-source datasets were adopted. The Classification and Recognition dataset includes 6,358 manually labeled samples spanning ten distinct traffic sign categories. The Vehicle and Pedestrian Detection dataset contains 5,748 images representing 8 different traffic objects such as pedestrians, small lorries, and trucks. The Pedestrian Augmented Traffic Light dataset consists of 2,397 traffic lights images labeld as "green", "red", "yellow", and "negative". The DVS-Gesture dataset is recorded using a DVS, consisting of spike trains with two channels corresponding to ON- and OFF event spikes. The dataset contains 11 hand gestures from 29 subjects under 3 illumination conditions, with 1176 training samples and 288 testing samples.

# Supplementary file

## 1. Supplementary Table 1

| Opcode | Operand | | | | | |
|---|---|---|---|---|---|---|
| FP_CONV | s_h_size | s_w_size | k_size | padding | stride | psum_acc |
| | c_size | m_size | c_offset | m_offset | Reserved | |
| | Compute convolution at FP ARRAY, fetch data from $s^{l-1}$ and $w^l$ from SRAM, refresh $Conv\_FP^l$ SRAM | | | | | |
| BP_CONV | du_h_size | du_w_size | k_size | padding | insert | psum_acc |
| | m_size | c_size | m_offset | c_offset | Reserved | |
| | Compute convolution at BP ARRAY, fetch data from $\nabla u^{l+1}$ and $w^{l+1}$ from SRAM, refresh $Conv\_BP^l$ SRAM | | | | | |
| WG_CONV | s_h_size | s_w_size | du_h_size | du_w_size | Insert | dw_acc |
| | dw_c_size | dw_m_size | dw_c_offset | dw_m_offset | Reserved | |
| | Compute convolution at WG ARRAY, fetch data from $s^l$ and $\nabla u^{l+1}$ from SRAM, refresh $\nabla w^{l+1}$ SRAM | | | | | |
| FP_SOMA | h_size | w_size | m_size | m_offset | pooling | t_acc |
| | Compute FP Soma, fetch data from $Conv\_FP^l$、$u^l$、Spike from SRAM, refresh $s^l$ SRAM | | | | | |
| BP_GRAD | h_size | w_size | c_size | c_offset | pooling | t_acc |
| | Compute BP Grad, fetch data from $Conv\_BP^l$、$\nabla u^l$、$u^l$、$s^l$、$fire'$ from SRAM, refresh $\nabla u^l$ SRAM | | | | | |
| FP_BN | h_size | w_size | m_size | m_offset | acc | Reserved |
| | Compute BN at FP Engine, refresh $Conv\_FP^l$ SRAM and BN private buffer | | | | | |
| BP_BN | h_size | w_size | m_size | m_offset | acc | Reserved |
| | Compute BN at BP Engine, refresh $\nabla u^{l+1}$ SRAM and BN private buffer | | | | | |
| FP_VECTOR | op_type | data_type | h_size | w_size | m_size | m_offset |
| | Compute VECTOR add/multiply/compare at FP Soma, refresh SRAM according to data_type, i.e. $Conv\_FP^l$、$s^l$ SRAM | | | | | |
| BP_VECTOR | op_type | data_type | h_size | w_size | m_size | m_offset |
| | Compute VECTOR add/multiply/compare at BP Grad, refresh SRAM according to data_type, i.e. $Conv\_BP^l$、$fire'$ SRAM | | | | | |
| NOC_DATA | flow_type | data_type | tag_id | Reserved | | |
| | Send a message to NOC, fetch data from SRAM | | | | | |
| NOC_CTRL | flow_type | tag_id | msg_box | | | |
| | Send a control message to NOC, trig an event at destination | | | | | |
| DMA_WR | data_type | src_addr | dst_addr | length | Reserved | |
| | Move data from sram to dram | | | | | |
| DMA_RD | data_type | src_addr | dst_addr | length | Reserved | |
| | Move data from dram to sram | | | | | |
| BARRIER | sub_type | Reserved | | | | |
| | BARRIER for instruction group | | | | | |

## 2. Supplementary Table 2

**Parameters of SNN**

| Parameters | Description |
|---|---|
| $N$ | batch size |
| $T$ | time steps of SNN |
| $H/W$ | input feature map height/width |
| $C$ | input feature map channel |
| $R/S$ | kernel height/width |
| $M$ | output feature map channel |
| $E/F$ | output feature map height/width |

## 3. Supplementary Table 3

### SRAMs of a Computing Core

| FP-Subcore | | BP-Subcore | |
|---|---|---|---|
| $s^{l-1}$ | 32KB | $\nabla u^{l+1}$ | 128KB |
| $w^l$ | 576KB | $w^{l+1\prime}$ | 576KB |
| $Conv\_FP^l$ | 128KB | $Conv\_BP^l$ | 128KB |
| $u^l$ | 128KB | $\nabla u^l$ | 128KB |
| $s^l$ | 32KB | $u^l$ | 128KB |
| Spike | 32KB | $s^l$ | 8KB |
| Others | 16KB | $\nabla w^{l+1}$ | 9 KB |
| | | Others | 24KB |
| Total | | 2073KB | |

## 4. Supplementary Table 4

### Hardware Utilization on VCU128 FPGA

| Component | LUT | FF | BRAM | URAM | DSP |
|---|---|---|---|---|---|
| FP engine | 208K (16%) | 176K (6.7%) | 48 (2.4%) | 192 (20%) | 1284 (14.2%) |
| BP engine | 268K (20.5%) | 248K (9.5%) | 36 (1.8%) | 224 (23.3%) | 2400 (26.6%) |
| WG engine | 184K (14.1%) | 132K (5.1%) | 0 | 0 | 1048 (11.6%) |
| Controller | 8K (0.6%) | 8K (0.3%) | 512 (25.4%) | 0 | 0 |
| Interconnect | 289K (22.2%) | 395K (15.2%) | 519 (25.7%) | 0 | 0 |
| Total used | 957K (73.4%) | 959K (36.8%) | 1115 (55.3%) | 416 (43.3%) | 4732 (52.4%) |

# 5. Supplementary Figure 1

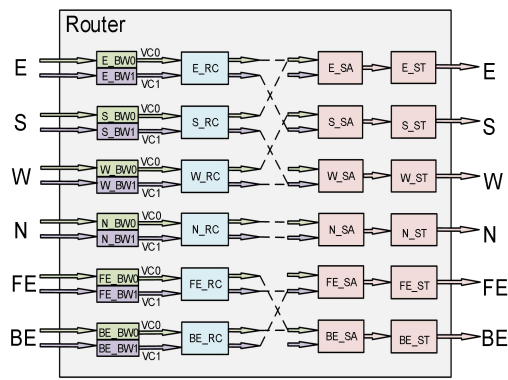 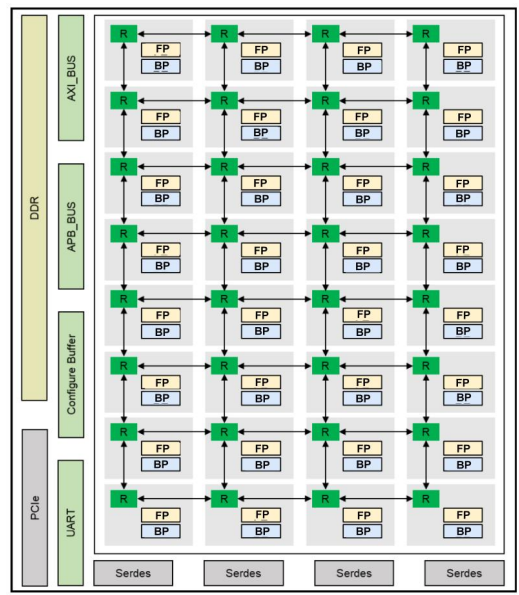

**The multi-core neuromorphic architecture.** a) Router module. b) A 32-core architecture.

# 6. Supplementary Figure 2

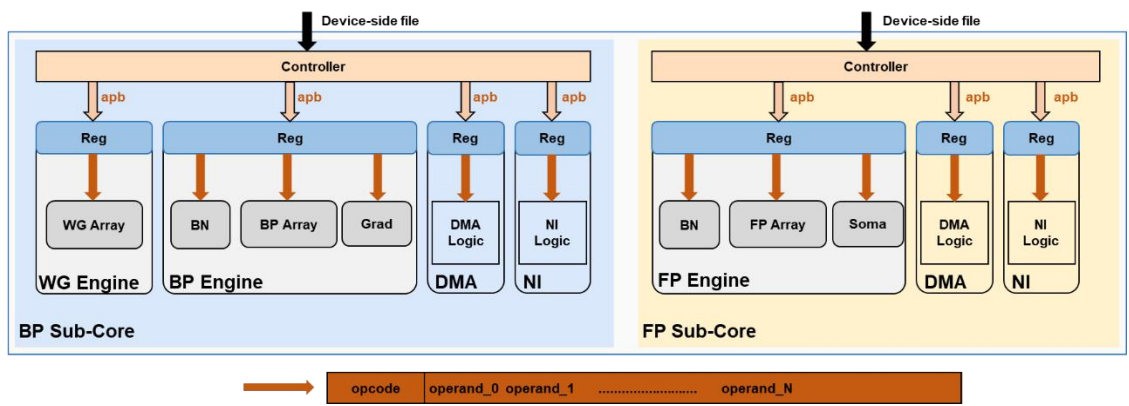

**The Controller and instruction generation in the multi-core neuromorphic architecture.** The Controller unit in each sub-core parses the device-side file generated by model compilation, generating instructions for modules in the sub-core. It monitors the working status of the sub-core's modules, and coordinated the orderly operation of all modules.

# 7. Supplementary Figure 3

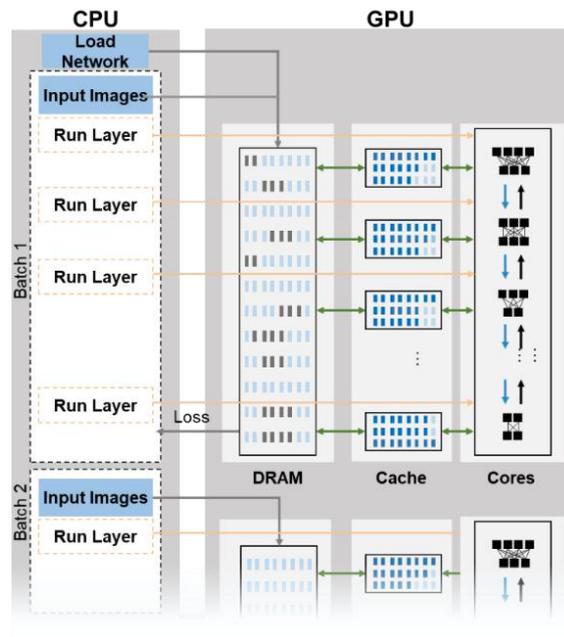

**The training procedure in GPU system.**

# 8. Supplementary Figure 4

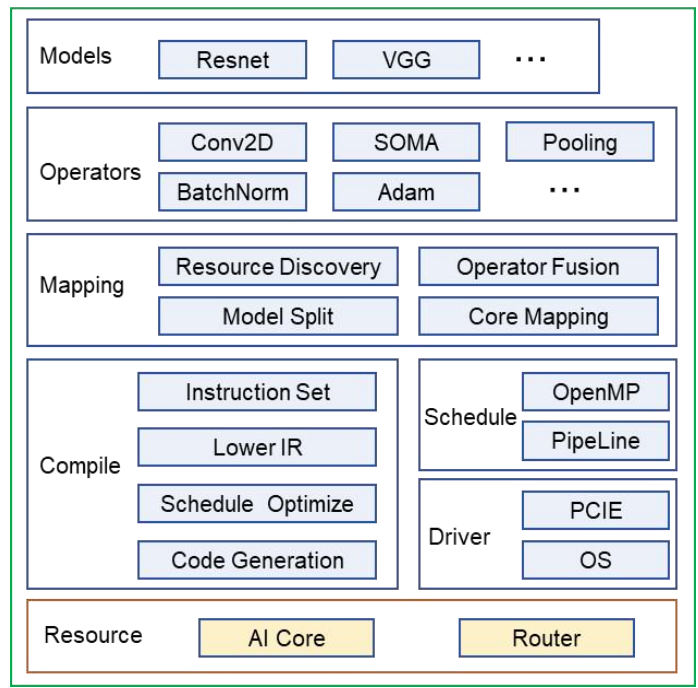

**Software toolchain for hardware implementation of SNN training.**

## 9. Supplementary Figure 5

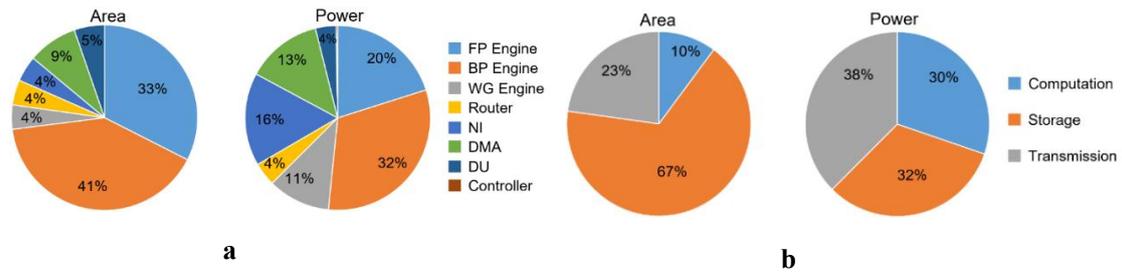

**Area and power estimation based on RTL simulation.** (a) Estimation of area and power consumption of modules in a computing core. (b) Area and power breakdown of computation, storage, and transmission.